\documentclass[twocolumn,floatfix,preprintnumbers,citeautoscript,newabstract,prl,superscriptaddress]{revtex4} 
\usepackage{amsmath,amssymb,graphics}
\usepackage[DIV12]{typearea}
\usepackage{amsmath,amsfonts,amssymb}
\usepackage{graphicx}
\usepackage{units}
\usepackage{mathrsfs}
\usepackage{bbm}
\usepackage{pifont}
\usepackage{braket}
\usepackage{units}
\usepackage[small,loose]{subfigure}  
\usepackage[dvipsnames]{xcolor}
\usepackage{mathtools}
\usepackage{tikz}
\usepackage[bookmarks=false]{hyperref}
\usepackage{enumerate}
\usepackage[normalem]{ulem}
\usepackage{cancel}
\usepackage{pdfcomment}

\newcommand{\eVdist}{\kern-0.06em}
\newcommand{\Ev}{\:\text{e\eVdist V}}     
\newcommand{\Kev}{\:\text{ke\eVdist V}}
\newcommand{\Mev}{\:\text{Me\eVdist V}}
\newcommand{\Gev}{\:\text{Ge\eVdist V}}

\newcommand{\ev}{\:\text{e\eVdist V}}   

\newcommand{\cm}{\:\text{cm}}

\newcommand{\ChargeC}{\ensuremath{\mathcal{C}}}

\newcommand{\TRH}{\ensuremath{T_{\mathrm{R}}}}
\newcommand{\FermiEnergy}{\ensuremath{\varepsilon_\mathrm{F}}}
\newcommand{\CnuBth}{\ensuremath{\text{C}\nu\text{B}_\mathrm{th}}}

\allowdisplaybreaks[1]

\def\mytitle{Non--thermal cosmic neutrino background}

\title{\boldmath\mytitle\unboldmath}

\begin{document}

\preprint{UCI-TR-2015-13; TUM-HEP 1009/15; FLAVOR-EU 106\\}

\title{Non--thermal cosmic neutrino background}

\author{Mu--Chun Chen}
\email[]{muchunc@uci.edu}
\affiliation{Department of Physics and Astronomy, University of California, Irvine, California 92697--4575, USA} 

\author{Michael Ratz}
\email[]{michael.ratz@tum.de}
\affiliation{Physik--Department T30, Technische Universit\"at M\"unchen, James--Franck--Stra\ss e~1, 85748 Garching, Germany
}
\author{Andreas Trautner}
\email[]{andreas.trautner@tum.de}
\affiliation{Physik--Department T30, Technische Universit\"at M\"unchen, James--Franck--Stra\ss e~1, 85748 Garching, Germany
}
\affiliation{Excellence Cluster Universe, Boltzmannstra\ss e~2, 85748 Garching, Germany
}

\date{September 2, 2015}

\begin{abstract}
We point out that, for Dirac neutrinos, in addition to the standard thermal
cosmic neutrino  background (C$\nu$B) there could also exist a non--thermal
neutrino background  with comparable number density. As the right--handed
components are essentially decoupled from the thermal bath of standard model
particles, relic neutrinos with a non--thermal distribution may exist until
today. The relic density of the non--thermal (nt) background can be constrained
by the usual observational bounds on the effective number of massless degrees of
freedom $N_\mathrm{eff}$, and can be as large as
\mbox{$n_{\nu_{\mathrm{nt}}}\lesssim 0.5\,n_\gamma$}. In particular,
$N_\mathrm{eff}$ can be larger than 3.046 in the absence of any exotic states.
Non--thermal relic neutrinos constitute an irreducible contribution to the
detection of the C$\nu$B,  and, hence, may be discovered by future experiments
such as PTOLEMY.  We also present a scenario of chaotic inflation in which a 
non--thermal background can naturally be generated by inflationary preheating.
The non--thermal relic neutrinos, thus, may constitute a novel window into the
very early universe.
\end{abstract}

\maketitle

\section{Introduction}

It is well established by the observation of neutrino oscillations  that at
least two neutrino species have non--vanishing mass. In the standard picture of
early universe cosmology, an upper bound on the sum of the neutrino masses can
be inferred \cite{Ade:2015xua},
\begin{equation}\label{eq:UpperBound}
 \sum m_{\nu} ~<~ 0.23\ev ~~\text{at 95\% CL}\;.
\end{equation}
Arguably, the most straightforward way to reconcile neutrino masses with the
standard model (SM) is to introduce three right--handed (RH) ``neutrino'' Weyl
spinors $\nu_\mathrm{R}^j$ ($1\le j\le3$) which are neutral under
$G_{\mathrm{SM}}$ and only couple to the SM via Yukawa interactions of the form
\begin{equation}\label{eq:YukawaCoupling}
 \mathscr{L}_\nu~=~Y_\nu^{ij}\,\overline{\ell}^i_\mathrm{L}\cdot \widetilde{H}\,\nu_\mathrm{R}^j
  +\text{h.c.}\;.
\end{equation}
Here $Y_\nu$ denotes the matrix of Yukawa couplings, $\ell_\mathrm{L}^i$ are the
left--handed (LH) lepton doublets and $H$ is the Higgs doublet. After
electroweak symmetry breaking, left-- and right--handed Weyl spinors combine to 
a massive Dirac fermion. The Dirac neutrino masses are given by the singular
values of $Y_\nu$ times the Higgs vacuum expectation value. In order to get
masses of $\mathcal{O}(0.1)\Ev$ or below, the eigenvalues of $Y_\nu$ need to be
$10^{-12}$ or smaller. A possible (lepton number violating) Majorana mass term
$M_{\cancel{L}}\, \overline{\nu_\mathrm{R}^\ChargeC}\nu_\mathrm{R}$ may be
forbidden by  a discrete subgroup of a baryon--minus--lepton number ($B\!-\!L$)
symmetry, or by other means. Throughout this work we will assume that there are no
Majorana mass terms.

The smallness of the Yukawa couplings and the absence of other interactions
implies that RH neutrinos are essentially decoupled from the
thermal bath in the early universe \cite{Antonelli:1981eg}.
As is well known (cf.\ e.g.\ \cite{Kolb:1990vq}), this means that RH neutrinos are not created in
any significant number from the thermal bath. Another consequence, however, is
that any previously existing abundance of RH neutrinos would not
thermalize. RH neutrinos would only be affected by a
red--shifting of their kinetic energy as well as a dilution of their number
density due to the Hubble expansion of the universe.

In the early universe, neutrinos are highly relativistic. Therefore, left-- and
right--handed components of the Dirac spinor  do not mix significantly  and can
be thought of as individual species.  Long after the decoupling of LH neutrinos
from the thermal bath neutrinos become non--relativistic, and thus form a
massive Dirac fermion. We refer to this process as ``left--right
equilibration''. For the originally thermally coupled LH neutrinos this implies
that half of them are converted to the RH component, thus, halving the
experimental count rate \cite{Long:2014zva} in neutrino capture experiments. 
For the originally decoupled RH neutrinos, on the other hand, this implies that
half of them are converted to the LH component, thus allowing for their
experimental detection through weak currents. The distinction between  left--
and right--handed neutrinos, therefore, is obsolete when the neutrinos are
non--relativistic. We will refer to the different  massive Dirac neutrinos as
``thermal'' and ``non--thermal'' neutrinos, according to their origin.

Let us briefly recall the usual ``standard'' scenario of a decoupled
relativistic species with a thermal spectrum. Here, the crucial assumption is
that the species has been in thermal equilibrium with the bath at some earlier
epoch. At decoupling, the decoupled species exhibits an essentially thermal
spectrum with a characteristic energy scale, i.e.\ the expectation value of the
particle energy $\langle E \rangle$, given by the temperature of the bath at
decoupling $\langle E \rangle\sim T_{\mathrm{dec}}$. After decoupling, the
decoupled species still obeys a thermal spectrum with a characteristic energy
scale given by the red--shifted decoupling temperature. This temperature will be
lower than the temperature of the non--decoupled components of the bath by a
factor of $(g_{*\mathrm{S}}(T_{\mathrm{today}})/g_{*\mathrm
S}(T_{\mathrm{dec}}))^{1/3}$, with $g_{*\mathrm{S}}$ being the effective number 
of degrees of freedom in entropy, due to entropy conservation. Thus, if the
decoupling happened sufficiently early, the number density of decoupled relics
gets diluted below any detectable value.  SM LH neutrinos have a relatively low
decoupling temperature of $T\sim 1\,\Mev$ from which an average number density
of \mbox{$n_{\nu_\mathrm{L}}\sim 336\,\cm^{-3}$} today can be anticipated. This is the
well--known thermal cosmic neutrino background (C$\nu$B)\, which should be
compared to the cosmic microwave background (CMB) of photons with an average
relic abundance of \mbox{$n_\gamma=410\,\mathrm{cm}^{-3}$}. It will be interesting to
see how the usual local density of neutrinos \cite{Ringwald:2004np} changes in
the presence of the additional, non--thermal degrees of freedom.

RH neutrinos may get populated through some beyond--SM interactions.
Examples include scenarios with a gauged $B\!-\!L$ symmetry which is broken at
the TeV
scale~\cite{Anchordoqui:2011nh,Anchordoqui:2012qu,SolagurenBeascoa:2012cz}. Here
one obtains a RH neutrino background with an almost thermal spectrum that
is, at the time of BBN, colder than the SM particles. In case 
the thermal RH neutrinos decouple at temperatures above the electroweak phase transition, 
their relic number density is smaller than \mbox{$\sim 36\,\mathrm{cm}^{-3}$}.

Let us now consider the case in which a species has never been in thermal
equilibrium with the rest of the universe.  Then the
spectrum of the decoupled species can be non--thermal. The energy of the
individual quanta, of course, still gets red--shifted and
their number density is still diluted by the Hubble expansion. Nevertheless, the
typical energy scale of the spectrum  is no longer required to be anywhere close
to the temperature of the bath at the time the decoupled particles are created.
This leaves some room to raise $\langle E \rangle$ above $T$, such that the
subsequent ``reheating'' of the bath can be compensated, and one is left with a
non--negligible relic density of the decoupled species. How much the relic
density can be raised crucially depends on how much the total energy density of
the universe can be raised. In standard cosmology, the energy density is
constrained by observational bounds on the universe expansion rate during the
epoch of big bang nucleosynthesis (BBN) and during the formation of the CMB
(cf.\ \cite{Dolgov:2002wy,Lesgourgues:2006nd} for reviews). The constraints are
typically quoted in terms of the number of effective relativistic
degrees of freedom $N_{\mathrm{eff}}$ that are thermally coupled to the bath
corresponding to an energy density
\begin{equation}\label{eq:extra_energy}
 \rho~=~N_{\mathrm{eff}}\,\frac{7}{8}\left(\frac{4}{11}\right)^{4/3}\,\rho_\gamma\;.
\end{equation}
The current bounds still allow for an additional energy density corresponding to
\cite{Ade:2015xua, Mangano:2005cc}
\begin{equation}\label{eq:NeffBounds}
 \Delta N_{\mathrm{eff}}~=~N_{\mathrm{eff}}-3.046~=~0.2\pm0.5 ~~\text{at 95\% CL}\;.
\end{equation}
Translated into a bound on the energy density of decoupled non--thermal RH neutrinos, 
we will show that a relic number density as high as \mbox{$n_{\nu_\mathrm{nt}}\sim 217\, \cm^{-3}$} 
can be consistent with data. 

An important question is whether the scenario of an existing non--thermal
spectrum of RH neutrinos is well motivated. That is, whether there exists a
process in the early universe that could produce such a spectrum.
Interestingly, coherent oscillations of the inflaton field at the end of
inflation can very efficiently produce particles non--thermally via the
parametric resonance (cf.\ \cite{Traschen:1990sw}). 
This so--called ``preheating'' occurs rapidly and far from thermal equilibrium.
That is,  it occurs before the perturbative decay of the inflaton and, thus, 
naively at a scale which is above the reheating temperature.  We will see that
already the simplest scenarios of ``fermionic preheating'' \cite{Greene:2000ew}
can successfully produce non--thermal relic neutrinos. In this case,
non--thermal relic neutrinos would constitute a probe of the universe 
at the time of inflation.

\section{Non--equilibration of \texorpdfstring{$\boldsymbol{\nu_\mathrm{R}}$}{NuR}}
\label{sec:NonEquilibration}

It is well--known that RH neutrinos are not thermally produced
\cite{Antonelli:1981eg}. Let us briefly recall how this also implies that any
existing abundance of RH neutrinos is not thermalized.

The only interaction of RH neutrinos with SM particles is the Yukawa coupling
\eqref{eq:YukawaCoupling}.  The corresponding interaction rate at temperatures
above the one of the electroweak phase transition can be estimated to be
\begin{equation}\label{eq:RateAboveEW}
 \Gamma_{\mathrm{RH}}(T\gtrsim v_{\mathrm{EW}})~\sim~  |y_\nu|^2   \,T\;.
\end{equation} 
Comparing this to the expansion rate of the universe, $H~\sim~T^2/M_\mathrm{P}$,
we find that this interaction is effective only for temperatures
\begin{equation}\label{eq:EqTempAboveEW}
 T~\lesssim~10^{-24}\,M_\mathrm{P}~\sim~\Kev\;,
\end{equation}
which is far below the range of validity of \eqref{eq:RateAboveEW}.

After the electroweak phase transition, interaction rates of RH
neutrinos with the plasma are suppressed by a factor proportional to $(m_\nu/E)^2$ relative 
to the interaction rates of left handed neutrinos, $\Gamma_{\mathrm{LH}}\sim
G_\mathrm{F}^2\,T^5$. Therefore,
\begin{equation}\label{eq:RateBelowEW}
 \Gamma_{\mathrm{RH}}(T\lesssim v_{\mathrm{EW}})~\sim~G_\mathrm{F}^2\,m_\nu^2\,T^3\;.
\end{equation}
Comparing this to $H$, we find that this interaction is effective only for temperatures 
\begin{equation}\label{eq:EqTempBelowEW}
 T~\gtrsim~10^{12}\Gev\;,
\end{equation}
which is far above the range of validity of $\eqref{eq:RateBelowEW}$.

This shows that RH neutrinos do not equilibrate  
either above or below the electroweak phase transition.

\section{Relic abundance of \mbox{non--thermal \texorpdfstring{$\boldsymbol{\nu_\mathrm{R}}$}{NuR}}}

Let us assume that there is a population of non--thermal RH neutrinos in the
early universe. In general, the neutrinos may have a distribution $f$ which
depends non--trivially on the neutrino energy $\varepsilon$. In what follows, we
discuss the extreme case in which the RH neutrinos  form a degenerate Fermi gas
at (RH neutrino) temperature equal to zero. The RH neutrinos are completely
decoupled from the thermal bath, which we assume to have temperature $\TRH$ at
the time the RH neutrinos are produced. At this time, the number and energy
densities of the RH neutrinos are given by 
\begin{equation}\label{eq:FDdensities}
 n_{\nu_\mathrm{R}}~=~\frac{g}{6\pi^2}\,\FermiEnergy^3\;
 \quad\text{and}\quad
 \rho_{\nu_\mathrm{R}}~=~\frac{g}{8\pi^2}\,\FermiEnergy^4\;,
\end{equation}
respectively. Here $\FermiEnergy$ is the Fermi energy and $g$ counts
the degrees of freedom and is $2$ for a Weyl fermion. 

Other non--thermal distributions $f$ may be considered. We will also discuss a
non--degenerate Fermi gas in which not all states below $\FermiEnergy$ are
occupied. This is accomplished by introducing a ``filling factor'' $0\le\eta\le1$
multiplying both densities \eqref{eq:FDdensities}. In order keep our
presentation simple, we will set $\eta=1$ here and comment on the case
$\eta\neq1$ below.

Besides $g$ and $\eta$, the only free parameter of our scenario is
$\xi~:=~\FermiEnergy/\TRH$. In order not to spoil the picture of a standard
model radiation dominated cosmic evolution  we require that
$\rho_{\nu_\mathrm{R}}\ll\rho_{\mathrm{rad}}$. As we find that all values of 
$\xi$ which are consistent with observation always respect this requirement, there is no need to state the
corresponding bound explicitly.

In a radiation dominated universe entropy conservation implies that the scale
factor is proportional to $R\propto g_{*\mathrm{S}}^{-1/3}T^{-1}$.  One may
wonder whether entropy conservation is spoiled by the fact that there is a
non--zero chemical potential for the RH neutrinos. This is not the
case due to the fact that the RH neutrinos are 
essentially non--interacting implying that their particle number is conserved
and the standard form of entropy conservation is maintained. 

The scaling of the particle number and that of the energy density then are given by
\begin{equation}\label{eq:density}
n_{\nu_\mathrm{R}}(T)~=~\frac{g\,\xi^3}{6\,\pi^2}\frac{g_{*\mathrm
S}(T)}{g_{*\mathrm{S}}(\TRH)}\,T^3\;,
\end{equation}
and
\begin{equation}\label{eq:edensity}
\rho_{\nu_\mathrm{R}}(T)~=~\frac{g\,\xi^4}{8\,\pi^2}\left(\frac{g_{*\mathrm{S}}(T)}{g_{*\mathrm{S}}(\TRH)}\right)^{4/3}\,T^4\;.
\end{equation}

The relic density of non--thermal neutrinos today can be obtained from the
scaled density of the originally RH neutrinos and is given by
\begin{equation}
 \frac{n_{\nu_{\mathrm{nt}}}}{n_\gamma}
 ~=~\frac{n_{\nu_\mathrm{R}}(T_\gamma)}{n_\gamma}
 ~=~
 \frac{g}{12\,\zeta(3)}\,\frac{g_{*\mathrm{S}}(T_\gamma)}{g_{*\mathrm{S}}(\TRH)}\,\xi^3\;.
\end{equation}
It is straightforward to translate the observational bounds on the energy
density in the universe into a constraint on the energy density of the
degenerate RH neutrinos during BBN or after CMB formation. From
\eqref{eq:extra_energy} and \eqref{eq:edensity} we obtain at BBN
\begin{equation}
 \Delta N^{(\nu_\mathrm{R})}_{\mathrm{eff}}
 ~=~
 \frac{8}{7}\,\frac{30}{8\,\pi^4}\,\frac{g\,\xi^4}{2}\,
 \left(\frac{g_{*\mathrm{S}}(T_{\mathrm{BBN}})}{g_{*\mathrm{S}}(\TRH)}\right)^{4/3}\;.
\end{equation}
The current observational limit \eqref{eq:NeffBounds} implies $\Delta
N^{(\nu_\mathrm{R})}_{\mathrm{eff}}\lesssim0.7$. In case there are  three
generations of relic non--thermal neutrinos with equal $\xi$, we take $g=6$ and
find
\begin{equation}\label{eq:xibound}
 \xi~\lesssim~3.26\;,
\end{equation}
where we have used $g_{*\mathrm{S}}(T_{\mathrm{BBN}})=10.75$ and
$g_{*\mathrm{S}}(\TRH)=106.75$. This corresponds to a relic density of 
\begin{equation}\label{eq:relicDensity}
n_{\nu_{\mathrm{nt}}}~\lesssim~0.53\,n_\gamma~\approx~217\,\mathrm{cm}^{-3}\;.
\end{equation}
This should be compared to the abundance of \CnuBth\ thermal relic neutrinos
\mbox{$n_{\nu_\mathrm{th}}\sim 336 \cm^{-3}$}. 

In principle, there could also be different values of $\xi$ for different
generations of RH neutrinos. Irrespective of this assumption, however, the total
relic density is bounded  from above by \eqref{eq:relicDensity}.

Let us comment on how our results change if we allow for a non--trivial filling
factor $\eta$. Assuming that $\xi$ is equal for all generations,
the relic density can be written in the form  
\begin{equation}
\frac{n_{\nu_{\mathrm{nt}}}}{n_\gamma}~\approx~1.2\,\frac{g_{*\mathrm
S}(T_\gamma)}{g_{*\mathrm{S}}(T_{\mathrm{BBN}})}\,\eta^{1/4}\,g^{1/4}\,\left(\Delta N^{(\nu_\mathrm{R})}_{\mathrm{eff}}\right)^{3/4}\;.
\end{equation}
This allows us to determine the maximal relic density for some given value of $\eta$
while keeping $\Delta N^{(\nu_\mathrm{R})}_{\mathrm{eff}}$ at the observational
upper bound. We see that the relic density can be sizable even for low values of
$\eta$.

Note that this discussion also shows that it is possible to explain sizable
deviations in $N_{\mathrm{eff}}$ without introducing any exotic particles.

\section{A concrete scenario for \texorpdfstring{$\boldsymbol{\nu_\mathrm{R}}$}{NuR} production}

In what follows, we will discuss a scenario in which a non--thermal neutrino
background can be naturally generated. Let us stress that the main point of this
Letter does not rely on this specific possibility,  which is just an existence
proof of a scenario with the desired properties. In the simplest cases of
``fermionic preheating'' \cite{Greene:1998nh} the inflaton $\phi$ is assumed to
have a potential  $V(\phi)=\frac{1}{2}m^2_\phi\phi^2$ and a Yukawa coupling
$\lambda\,\phi\,\overline{\Psi}\Psi$, where $\Psi$ is a Dirac fermion.
Preheating via the parametric resonance \cite{Kofman:1994rk} then produces
fermions with a non--degenerate Fermi spectrum, i.e.\ with momenta
stochastically filling a sphere of radius $\FermiEnergy\sim q^{1/4}m_\phi$, 
where $q:=\lambda^2\phi_0^2/m^2_\phi$ is the so--called resonance parameter and
$\phi_0$ the initial displacement of $\phi$.  After a couple of inflaton
oscillations this process becomes ineffective and the inflaton decays
perturbatively seeding the hot early universe. In addition to the perturbative
decay, reheating could also occur via a coupling $\phi^2H^2$ to the Higgs portal
and the scalar preheating mechanism \cite{Kofman:1994rk}, thus producing the
usual SM particles.

Since $q$ a priori is a free parameter it is possible to obtain a characteristic
energy of the non--thermal spectrum $\langle E\rangle\sim \FermiEnergy$  which
can be much bigger than the naive reheating temperature of $\TRH\sim m_\phi/2$.
Thus, a non--thermal spectrum may be created with a number density which is
non--negligible even today.

A possible scenario for the production of a non--thermal neutrino background
is, therefore, based on the coupling
\begin{equation}\label{eq:CouplingInflatonRNneutrino}
 \mathscr{L}
 ~\supset~
 \lambda\,\phi\,\overline{\nu_\mathrm{R}^\ChargeC}\nu_\mathrm{R}
 +\text{h.c.}\;.
\end{equation}
For this coupling to be allowed the inflaton must be appropriately charged under
the symmetry that prohibits the Majorana mass term of $\nu_\mathrm{R}$. This
also implies that the vacuum expectation value of $\phi$ must vanish
such that the Majorana mass term is not reintroduced.

The non--thermal neutrino spectrum is then assumed to be created via
fermionic preheating directly after inflation and, thus, can be approximated by 
a non--degenerate Fermi--Dirac distribution at zero temperature.

For values of $\xi\lesssim3$ and with the naive reheating temperature given  by
$\TRH\sim m_\phi/2$ we find \mbox{$q\sim\xi^4\lesssim10^2$}. Even though there
has been no dedicated analysis in this direction, this value of $q$ seems to
easily allow for filling factors reaching $\eta\gtrsim0.3$ \cite{GarciaBellido:2000dc}.

\section{Detection of a non--thermal neutrino background}

The scaling of the density \eqref{eq:density} implies that the Fermi energy
scales linearly with $T$,
\begin{equation}
 \FermiEnergy(T)
 ~=~
 \left(\frac{g_{*\mathrm{S}}(T)}{g_{*\mathrm{S}}(\TRH)}\right)^{1/3}\xi\,T\;. 
\end{equation}
In particular, the characteristic energy of the non--thermal neutrinos is
$\langle E_{\nu_{\mathrm{nt}}}\rangle\sim\FermiEnergy$, and they are 
non--relativistic at late times, just as the standard \CnuBth.  Thus, after 
left--right equilibration  half of the initially RH neutrinos may be detected
via an inverse beta--decay in neutrino capture experiments such as PTOLEMY
\cite{Betts:2013uya}. Even though the projected energy resolution could resolve 
an ``electron neutrino mass'' close to its upper bound \eqref{eq:UpperBound}, 
it will not suffice to resolve the spectrum of relic neutrinos. For this reason
the non--thermal neutrinos constitute an irreducible contribution to any planned
experiment which is sensitive to the C$\nu$B. Far future experiments  with a
substantially improved neutrino energy resolution, however,  could distinguish
the contributions of thermal and non--thermal neutrinos.

Since the thermal neutrinos propagate as mass eigenstates, the different flavors
will,  to a good approximation, be equilibrated at late times
\cite{Long:2014zva}. We further assume that the flavor composition of
the non--thermal neutrinos is also roughly $1\,:\,1\,:\,1$.  The maximal global
number density of relic  non--thermal neutrinos which is available for
electron--neutrino capture is then given  by \mbox{$\sim 36\,\mathrm{cm}^{-3}$}.
Comparing this with the  detectable number density of \CnuBth\ neutrinos which is
$\sim56\,\mathrm{cm}^{-3}$,  we see that any experiment which aims for detecting
the C$\nu$B should be able to detect the non--thermal Dirac neutrino
background.

Recently it has been suggested that measurements of the relic neutrino abundance
could discriminate between Dirac and Majorana neutrinos  due to their different
projected count--rates for PTOLEMY of $\sim4\,\mathrm{yr}^{-1}$ and
$\sim8\,\mathrm{yr}^{-1}$, respectively \cite{Long:2014zva}. We see that this
proposal may not work in the presence  of additional non--thermal Dirac
neutrinos which could increase the respective count rate by 64\% thereby
diminishing the difference between Dirac and Majorana neutrinos. 

\section{Summary}

\begin{figure}[ht]
\includegraphics[width=1.0\linewidth]{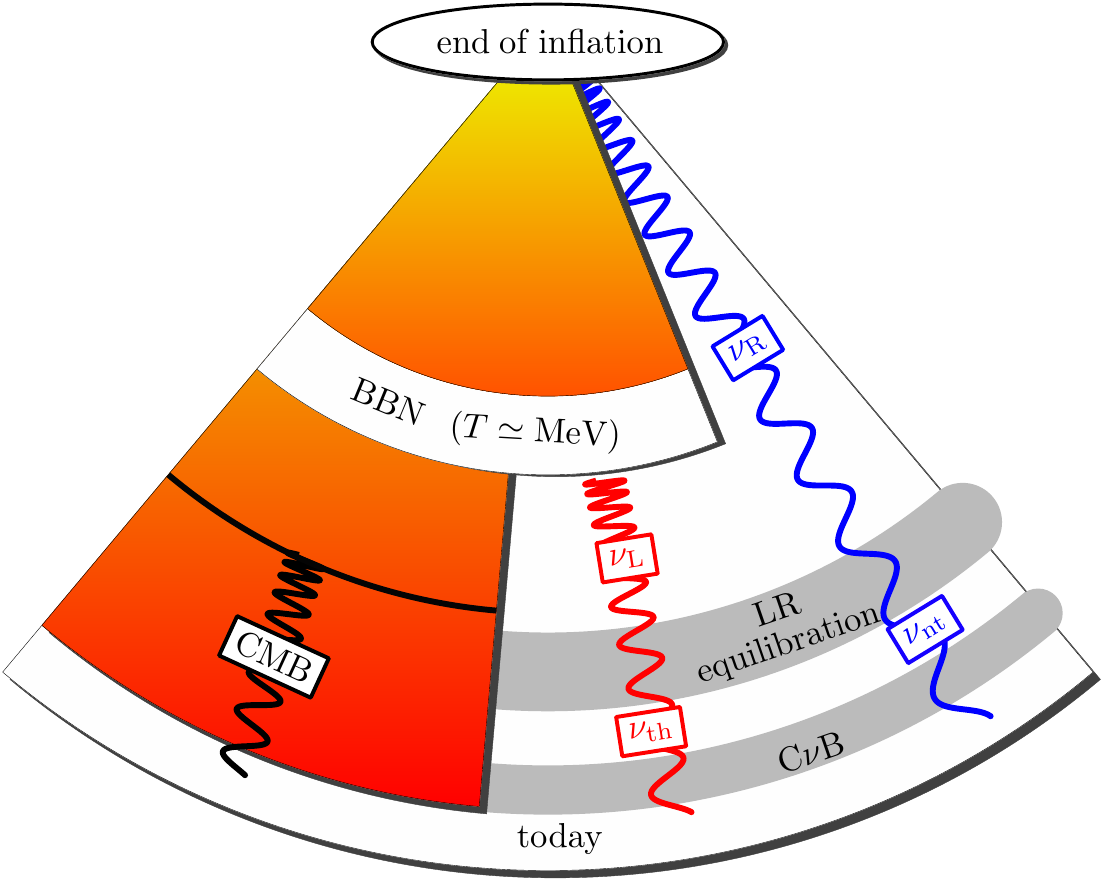}
\caption{Cartoon of the history of the universe with non--thermal neutrinos.}
\label{fig:NonThermal1}
\end{figure}

\begin{figure}[ht]
\includegraphics{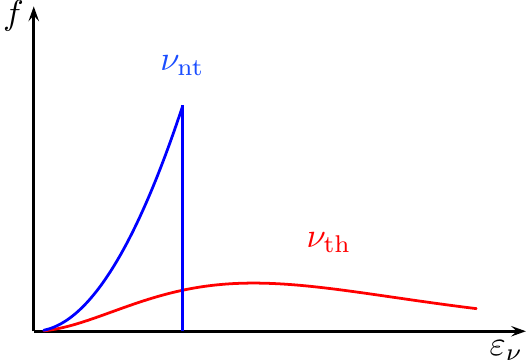}
\caption{Phase space distributions of thermal and non--thermal neutrinos during
BBN.}
\label{fig:NonThermal2}
\end{figure}

We depict the basic points of our scenario in Figure~\ref{fig:NonThermal1}. In
the very early universe, a significant number of $\nu_\mathrm{R}$ states are
created with a non--thermal spectrum. They are decoupled until very late  and
have during BBN an average energy below the one of the thermal neutrinos
(Figure~\ref{fig:NonThermal2}). This leads to a situation where the contribution
of non--thermal neutrinos to the energy density of the universe is consistent
with observation. Yet the relic abundance of non--thermal neutrinos can be as
large as \mbox{$\sim0.5\,n_\gamma$} today. This has important implications for
the prospects of discovering the C$\nu$B as well as for the clustering of relic
neutrinos. Note that our scenario can explain deviations of
$N_\mathrm{eff}$ from its usual value 3.046 without the need to add any
extra states to the SM apart from right--handed neutrinos and the inflaton,
which is an ingredient of almost any realistic cosmology.

The CMB provides us with information on the universe at the time of photon
decoupling, which happened around $380,000$ years after the big bang.
Likewise, the thermal neutrino spectrum is sensitive to how the universe looked
like at the time of BBN, when it was roughly 1\,second old. This is to be
contrasted with what one could learn from the non--thermal neutrino background.
As illustrated in the discussion of the preheating scenario, a future
possible detection and subsequent careful examination of the non--thermal
neutrino background may provide us with a possibility to directly probe features
of inflation (Figure~\ref{fig:NonThermal1}). Assuming an inflation scale of the
order $10^{16}\Gev$, non--thermally produced right--handed neutrinos may allow
us to probe the universe when it was as young as $10^{-38}\,$seconds. The
data gained this way will be complimentary to what one can learn from
gravitational waves and a non--trivial tensor--to--scalar ratio.


\begin{acknowledgments}
M.-C.C.\ would like to thank TU M\"unchen for
hospitality. M.R.\ would like to thank the  UC Irvine, where part of this work
was done, for  hospitality. This work was partially supported by the DFG
cluster  of excellence ``Origin and Structure of the Universe''. 
The work of M.-C.C.\ was supported, in
part, by the U.S.\ National Science  Foundation under Grant No.\ PHY-0970173. 
M.-C.C.\ and  M.R.\  would like to thank the Aspen Center for Physics for hospitality and support.
This research was done in the context of the ERC Advanced Grant project
``FLAVOUR''~(267104).
\end{acknowledgments}

\bibliography{Orbifold}
\bibliographystyle{apsrev}
\end{document}